\documentclass[useAMS,usenatbib,usegraphicx]{mn2e}

\begin{document}

\title[Relic Radio Bubbles and Cluster Cooling Flows]
  {Relic Radio "Bubbles" and Cluster Cooling Flows} 
\author[D. S. De Young]
  {David S. De Young$^1$ \\
  $^1$ National Optical Astronomy Observatory, Tucson, AZ, USA}
\date{Accepted for publication - MNRAS}

\pagerange{\pageref{firstpage}--\pageref{lastpage}} \pubyear{2002}

\def\LaTeX{L\kern-.36em\raise.3ex\hbox{a}\kern-.15em
    T\kern-.1667em\lower.7ex\hbox{E}\kern-.125emX}

\newtheorem{theorem}{Theorem}[section]
\label{firstpage}

\maketitle

\begin{abstract}
Recent suggestions that buoyant radio emitting cavities in the
intracluster medium can cause significant reheating of cooling flows
are re-examined when the effects of the intracluster magnetic field
are included.  Expansion of the cavity creates a tangential field in
the ICM surrounding the cavity, and this field can suppress
instabilities that mix the ICM and the radio source.  Onset of
instability can be delayed for $\sim 10^{8}$ yr, and calculation of
the subsequent turbulent cascade shows that actual reheating of the
ICM may be delayed for up to $\sim 5 \times 10^{8}$ yr.  These
results may explain why the relic radio cavities remain as intact
entities at times $\ge 10^{8}$ yr, and the delay in injection of
energy from the radio source into the intracluster medium may
mean that the role of radio sources in reheating cooling flows
should be re-examined.  In addition, the existence of
relic radio cavities may also imply that the particle content of
radio source lobes is primarily electrons and protons rather than
electrons and positrons.
\end{abstract}


\section{Introduction}
The interaction of extended radio sources in clusters of galaxies with
the intracluster medium (ICM) has been inferred since the first
observations of "head-tail" radio sources over two decades ago.
However, direct observation of this interaction required imaging of
the ICM itself, and early hints of this came from the 
Einstein Observatory x-ray observations of Cygnus A (Harris et al. 1994).
Unmistakable evidence for interaction came from the $ROSAT$
high resolution images of NGC 1275 in the Perseus cluster
(B\"{o}hringer et al. 1993), and more 
recent high resolution x-ray observations with the Chandra
Observatory have provided clear and dramatic evidence for the 
interaction of radio sources with the hot intracluster medium
(e.g., Fabian et al. 2000).
These images, when coupled with high resolution radio observations,
clearly show the extended radio source lobes displacing the x-ray
emitting gas and forming a cavity in the ICM.  In general these
cavities show a rather regular and spheroidal
morphology, suggesting the presence
of a "relaxed" structure that is in approximate equilibrium
with its surroundings.  There
is no evidence for a high temperature layer surrounding the cavities
that would be produced by a shock front; in some cases the region of
bright x-ray emission surrounding cavity has a lower temperature
than the ambient ICM (Nulsen et al. 2002).
In addition, the x-ray observations of the cavity in A2597 
show that the pressure in the limb brightened region surrounding
the cavity is not higher than the ambient ICM pressure, but that
it is comparable to or somewhat lower than the ambient value 
(McNamara 2002a, McNamara et al. 2001).  The
densities in the limb brightened region are, however, higher than
that in the undisturbed ICM.  This is not what would be expected
if the surface of the cavity were expanding supersonically into
the ICM, since the resulting shock in front of the cavity surface
would yield a pressure significantly higher than ambient:
$p_{1}/p_{0} = (2 \gamma M_{0}^{2} - \gamma + 1)/(\gamma + 1)$,
where $M_{0} > 1$ is the Mach number of the shock relative to the
ambient gas, the subscripts $0$ and $1$ refer to pre- and post-
shock conditions, and $\gamma$ is the usual ratio of specific
heats.  The above relation is valid for arbitrary shock strengths
(e.g., De Young 2002). 
Moreover, the geometry of the cavities, and especially the "relic"
cavities discussed below, is not suggestive of the supersonic
expansion that would be expected for a radio lobe that is still
being supplied by collimated jets emanating from the nucleus of
the parent galaxy. As is well known, these energetic lobes have a
much more elongated geometry, such as that seen in the "classical"
FR-II radio sources.  The nearly spherical geometry of the x-ray
and radio morphologies under discussion here is consistent with
a radio lobe that has come into rough pressure equilibrium 
with the ambient medium.  Again, this is particularly true of the
relic cavities discussed below, where there is no clear connection
seen in radio emission between the relic and the active nucleus.
In addition, the inner radio source in A2597 shows a complex 
and contorted geometry that suggests that strong deceleration and
"frustration" may be taking place in the inner regions where the
extended source is still being fed by outflow from the nucleus.
This again would lead to prompt deceleration of the radio source
relative to the ICM, and when the outflow from the nucleus ceases,
such objects would rapidly come into pressure equilibrium with the
ambient gas. 
All of this is suggestive of rather slow, subsonic inflation rates for
the cavities, and if they are in pressure equilibrium with the
ambient ICM, then their internal energy densities exceed the 
minimum equipartition energy densities by roughly an order of
magnitude (McNamara et al. 2001). 
It is important to note that all the above arguments about the
expansion speed of the relics into the ICM are indirect.  Since no
expansion speeds can be measured, it is possible that the expansion
of these objects is still supersonic, in excess of $10^{8}$ cm s$^{-1}$,
in which case their internal energies are even higher than if they
were in pressure equilibrium.  In addressing this issue, 
Soker, Blanton \& Sarazin (2002) argue that certain combinations of
effects can result in a mildly supersonic expansion producing x-ray
signatures that are consistent with the data, though strong shocks
appear to be ruled out.
The consequences of possible 
supersonic expansion will be discussed subsequently. 

\subsection{The Influence Upon Cooling Flows}

The geometries and energy densities of the radio cavities naturally
suggest the occurrence of buoyant motion through the hot ICM in
the presence of the gravitational potential of the central galaxy
and cluster core.  The evolution of such buoyant cavities has been
suggested to be an important factor in the evolution of cluster
cooling flows. It has 
been known for many years that the total energies present in
extended radio sources in clusters ($10^{57}-10^{59}$ erg)
are enough to significantly
influence the overall energy budget of the ICM.  The problem has
been, and may still be, in finding an effective mechanism for
mixing this energy uniformly throughout the inner regions of the
ICM. Several authors have recently suggested that buoyant radio
cavities can accomplish this, either through advective mixing
of differing regions of the ICM at different temperatures or 
through the dissolution of the radio cavities and the dispersal
of their energetic and hot radio plasma into the ICM.
The two dimensional axisymmetric hydrodynamic calculations of Reynolds et al.
(2002) show some advection of cooler intracluster material 
into hotter regions of the ICM, though it is less
clear that a truly buoyant cavity  with a geometry similar to that
seen in the x-ray data is produced by 
these simulations. This may be due in part to the effects
of the axisymmetry constraint.
The three dimensional hydrodynamic calculations of Br\"{u}ggen et al.
(2002) show more clearly the development of a buoyant cavity, and
these simulations also show the development of Rayleigh-Taylor
instabilities that lead to mixing
with the ICM and eventual destruction of the cavity as a separate entity.
This process could in principle inject a significant amount of energy
into the ICM and can be more effective for reheating than
the mixing of different regions of the ICM via advection, though this
is also seen in the simulation of Br\"{u}ggen et al.  The two dimensional
very high resolution hydrodynamic simulations of Br\"{u}ggen and Kaiser
(2002) show very clearly the onset and non-linear development of 
surface instabilities in the rising cavity, and these elegant calculations
show that the cavity is destroyed and mixing well underway after the
cavity has risen only a few of its own diameters.  Presumably in
three dimensions this process could be even more effective due to the
larger number of modes available for the development of the instability,
(assuming that the resolution of the numerical simulation is
unchanged).  
Thus hydrodynamic consideration of buoyant cavities in the ICM show
that very significant mixing of the cavity material with the ambient
ICM can occur on short timescales, of order $10^{7}$ years,
and this can transfer a significant amount of the
energy from the radio source plasma to the intracluster medium.  The
final mixing of this material throughout the ICM has yet to be 
calculated; Br\"{u}ggen \& Kaiser (2002) 
provide an estimate of what might be the final state
of the ICM after such mixing by averaging the energy input from the
buoyant plumes over azimuth.  This estimate shows that, if 
complete mixing can occur on a time short compared with inflow times,
the injection of energy from radio sources can be a significant factor
in reheating cooling flows and hence may be one solution to the
long standing cooling flow problem.

\section{ The Evolution of Relic Radio Bubbles}

Recently, new radio and x-ray observations have revealed an additional
feature in the intracluster medium of some rich clusters, and this
is the presence of pairs of what appear to be relic radio "bubbles"
that lie well outside the more luminous radio cavities in the inner
core of the ICM.  Spectacular examples of this phenomenon are found
in the Perseus (A426) and A2597 clusters, though other examples have 
also been found (McNamara 2002b).
These relic bubbles are coherent objects that are 
nearly spherical in appearance and have nearly featureless radio
emission at low frequencies but are not easily seen at high frequencies
(Fabian et al. 2002). They also appear to be in equipartition 
with the ambient ICM at pressures of $1-4 \times 10^{-10}$ dyne cm$^{-2}$, 
and if so then they have internal
energy densities that are again in excess of the equipartition values
by factors of ten.  The buoyant rise times to their current positions
from the central galaxy are of order $10^8$ years, which exceeds the
radiative lifetimes of the electrons in the inner lobes by factors
of ten (McNamara  et al. 2001, Churazov et al. 2001).  This may imply the
need for electron re-energization (see also 
Br\"{u}ggen et al. 2002).

\subsection{The Role of Intracluster Magnetic Fields}

In light of the results from the numerical simulations,
one of the most surprising aspects of the relic radio bubbles is that
they are intact.  If such objects were to follow the evolutionary
path described by hydrodynamic simulations, one would expect that 
at distances $\sim 30$ kpc from the cluster center and at times
$\sim 10^8$ years the radio remnants would have become
fragmented and assimilated into the
ambient intracluster medium.  The fact that they have not done so
may have important implications for the overall energy balance in
the intracluster medium.  The reason for the preservation of the
relic radio bubbles may possibly be found in the magnetic fields that 
permeate the ICM.  It has been known for some time that the hot
gas in clusters of galaxies often has within it a significant
magnetic field (e.g., Carilli \& Taylor 2002, Taylor et al. 2002), with
typical magnetic field strengths of order $5 \times 10^{-6}$ G.  The
origins of such fields remain somewhat obscure, but their effects
on the evolution of radio sources in clusters
are significant.  As a radio source
near the cluster center begins to inflate a cavity in the hot ICM,
the ambient magnetic field is excluded from this cavity along with
the hot intracluster gas.  This will result in the external field
forming a sheath around the cavity in which the field is primarily
tangential to the cavity surface and has a higher value than the
ambient field due to the effects of compression, as can be seen from
simple flux conservation arguments.
As mentioned above, the x-ray data imply
that the cavity expands subsonically or transonically, and thus this
region of compressed and largely azimuthal field will expand into
the ambient ICM at the local signal speed, slightly ahead of the
advancing boundary of the inflating cavity.

This layer of tangential magnetic field around the
buoyant cavity will suppress the short wavelength and fastest
growing modes of the Rayleigh-Taylor (R-T) and Kelvin-Helmholtz (K-H)
instabilities.  Because the lifting and mixing of different
layers of the ICM is mediated primarily by these instabilities,
as is large-scale mixing of the energetic radio emitting plasma
with the ICM, the presence of this external layer of magnetic
field will have an important influence on the evolution of the
buoyant radio cavities.  A magnetic field parallel to the 
interface between two fluids of differing density will suppress
the R-T instability because the field acts as a source of surface
tension at the interface and thus stabilizes small wavelength
perturbations.  This stabilization occurs for all wavenumbers
larger than (e.g., Chandrasekhar 1961; Shore 1992)

\begin{equation}
k_{c} = \frac{2 \pi g (\rho_{2} - \rho_{1})}{B^{2} Cos^{2} \theta},
\end{equation}
where $g$ is the acceleration due to gravity, $\rho_{1}$ is the density
of the lighter fluid (the radio bubble) and $\theta$ is the angle between
the magnetic field vector and the wave vector of the perturbation.
For a tangled azimuthal field an average value 
of $<Cos^{2} \theta > = 1/2$
can be used.  The value of $k_c$ provides the wavenumber of marginal
stability, and formally the growth rate of this mode is zero.  
Perturbations of longer wavelength grow initially at a rate given
by 

\begin{equation}
n^{2} = {\frac{g k}{(\rho_{2}+\rho_{1})}}\left[(\rho_{2}-\rho_{1}) -
{\frac{k B^{2} cos^{2}\theta}{2 \pi g}}\right],
\end{equation}
where $n$ is the coefficient in the initial growth rate given by
$\exp {nt}$.  The wavenumber with the most rapid growth rate is found
by differentiating Eq. 2 with respect to $k$.  Note that the 
instability treated here is that of the surface of a buoyant bubble
rising in a gravitational field and not that of the surface of a
decelerating expansion.  It is assumed here that the radio relics
are in approximate pressure equilibrium with the ambient gas.  If
the relics are expanding supersonically then this will not be
the case, but the age of these relics and their disconnection from
the active nucleus, together with their geometry, strongly suggests
that they have come into approximate  
equilibrium with their surroundings.  The non-relic 
radio bubbles in the inner regions may be expanding at a mildly
supersonic rate, as mentioned in Introduction, and such supersonic
propagation speeds can act to suppress the R-T instability in those
objects.

A conservative estimate that produces the largest value of $k_c$
and the fastest growth rate is to assume that $\rho_{2} \gg \rho_{1}$.
For the relic bubbles in A2597 and Perseus the ambient number densities
are of order $10^{-2}$ cm$^{-3}$ at distances of $\sim$ 30 kpc from the
cluster center, which is the appropriate distance for these relic radio
bubbles.  
Using the gravitational potential of a central galaxy
of mass $10^{12}$ solar masses and 
an average value of the magnetic field appropriate for 
cooling flow clusters of
$5 \times 10^{-6}$ G (e.g., Carilli \& Taylor 2002), which assumes no 
significant amplification of the field in the
compressed sheath surrounding the bubble and which will maximize the
value of $k_c$, one finds $k_{c} = 1.37 \times 10^{-22}$, or
$$
\lambda_{c} = 2 \pi/k_{c} = 15.2 kpc.
$$
This is the shortest wavelength for the onset of the R-T instability
under these conditions. 
The wavenumber of the fastest growing mode in the linear regime,
obtained via differentiation of Eq. 2, is $k_{*} = k_{c}/\sqrt{3}$,
and substitution of this into Eq. 2 gives the maximum growth
rate as
$$
\Gamma_{*} = 1/n_{*} = 4.2 \times 10^{7} yr.
$$
This is the time required for an initial perturbation of 
wavelength $\lambda_{*}$ to grow by a factor of $e$; subsequent
growth into the non-linear regime that will result in disruption
of the radio cavity will require times greater than this by at
least factors of two.

The interface between the buoyant cavity and the ambient ICM is
also subject to the Kelvin-Helmholtz (K-H) instability between two
fluids in relative motion.  Again, the effective surface tension
of the tangential ICM field that surrounds the cavity can
suppress the onset of this instability.  In the absence of a gravitational
restoring force perpendicular to the fluid interface, which is
the case here, the flow is
stable against the K-H instability if (e.g., Chandrasekhar 1961) 

\begin{equation}
(U_{1} -  U_{2})^{2} \leq \frac{B^{2} (\rho_{1} + \rho_{2})}{2 \pi
 \rho_{1} \rho_{2}},
\end{equation}
where $U_{1}$ and $U_{2}$ are the velocities of the two fluids 
(in this case $U_{2} = 0$ and $U_{1}=U_{rel}$),
$B$ is the average value of the tangential magnetic field, and the
densities $\rho$ have the same meaning as previously.  The 
Kelvin-Helmholtz instability is of less importance as an agent for mixing
the intracluster medium with itself and with the hot radio source
plasma than is the Rayleigh-Taylor instability.  
This is because the K-H instability
will, in its fully developed non-linear form, lead to a turbulent
mixing layer along the surface of the radio source cavity (e.g., De Young
2002).  This will influence a much smaller volume of the ICM than
will the R-T instability that leads to the destruction and complete
mixing of the radio source bubble with the ICM. The ratio of these
two volumes is of order $V_{K-H}/V_{R-T} \sim \Delta R/ R $, where $R$
is the radius of the bubble and $\Delta R \ll R$ is the thickness of
the turbulent mixing layer. 

The relative speed of the buoyantly rising bubble is
clearly subsonic; the three-dimensional simulations of Br\"{u}ggen
et al. (2002) show a relative speed of $\simeq 2.5 \times 10^{7}$ cm s$^{-1}$,
and their analytic estimate
of the terminal speed of a rising bubble yields a similar value.
(See also Churazov et al. 2001.) 
Equation 3 actually applies to the case where the same magnetic
field exists on both sides of the interface.  The stability criterion
basically states that if the relative speed of the two flows is less
than the Alfv\'{e}n speed, then the Kelvin-Helmholtz modes are stabilized,
since perturbations are then damped by the
fast moving Alfv\'{e}n waves moving along the interface.
For the case of the relic radio bubbles the magnetic field
is different inside and outside the interface, and hence the relevant
Alfv\'{e}n speed for the two different regions must be used.  In the
exterior, the 
same intracluster conditions as used
in calculating the onset of the Rayleigh-Taylor instability gives
stability against the K-H modes if
$U_{rel} \leq 1.5 \times 10^{7}$ cm s$^{-1}$.
For the interior the conditions are less well known; pressure equilibrium
plus the lack of x-ray emission in the Chandra band pass suggests the 
presence of a hot and rarefied gas, and if the 
interior number density is of order $10^{-4}$ (cf. Sect. 3) 
and the magnetic fields
are near the equipartition values of a few $\mu$G, then again stability
against K-H modes occurs for 
$U_{rel} \leq 3 \times 10^{7}$ cm s$^{-1}$.  Thus these speeds are
comparable to or greater than the relative speeds of the buoyantly
rising bubbles obtained from the numerical simulations, and the interface
is marginally stable against the Kelvin-Helmholtz instability. 
The growth rates of this instability are of order $n \approx k
U_{rel}$, and for any appreciable mixing to occur (and for any observable
deformation of the bubble interface) the wavelength of the instability
should be of order $1$ kpc or more.  This then gives growth times of
order $10^{7}$ years or more.

\subsection{Implications for Re-energization of Cooling Flows} 

Use of the buoyant speeds obtained from the simulations (Br\"{u}ggen et al.
2002, Churazov et al. 2001) gives lifetimes for the relic radio bubbles
in A2597 of $10^{8}$ years and $\sim 5 \times 10^{7}$ years for those
in A426.  The three dimensional hydrodynamic 
simulations of Br\"{u}ggen et al. (2002)
and the similar high resolution two dimensional simulations of Br\"{u}ggen \&
Kaiser (2002), which are of the appropriate scale for
these objects, show that the rising cavities are clearly being disrupted
at times of $3 - 4 \times 10^{7}$ years at a distance 
from the nucleus of $\simeq 15$
kpc for the three dimensional case and at $\simeq 6 \times 10^{7}$ yr
and $\simeq 20$ kpc for the two dimensional calculation. (The two 
dimensional case has a higher energy input than is appropriate for
these cluster radio sources, and it also 
may be more stable due to suppression of some 3-D
modes.)  However, the relic radio bubbles are still intact and show
no signs of disruption at distances of $30$ kpc and ages of 
$10^{8}$ years.  Hence some additional processes other than purely
hydrodynamical ones must be at work, and the above calculations show
that the displaced intracluster magnetic field may provide the 
stabilizing influence that keeps the relic bubbles intact. 
In this case the 
Rayleigh-Taylor instability, which is the most disruptive, does not
even commence its growth in the small amplitude linear phase 
until times of about $5 \times 10^{7}$ years, and the
\emph{shortest} wavelength of the initial instability is $\simeq 15$
kpc, which is comparable to the radii of the relic bubbles in A2597
and is comparable to the overall size of the bubbles in A426.  This
is to be compared with the purely hydrodynamic simulations, which show
that the R-T instability has proceeded far beyond the linear scale at
times of $5 \times 10^{7}$ years, and at wavelengths that are much
less than the overall radius of the rising bubble. 
Hence both
the spatial and temporal scales for disruption of these relics 
in the presence of external magnetic fields are 
such that this instability has not caused disruption by their
current age.  This then implies a significant reduction of the mixing
of energetic radio source material with the ambient ICM, and in addition
it may imply a lowered efficiency in "lifting" one part of the ICM into
another.  This is because, in the purely hydrodynamic case, 
the "lifting" is due to boundary layer
(as opposed to mixing layer) effects, and the unperturbed surface of 
the bubble has both a thinner boundary layer than a surface roughened
by small wavelength instabilities and a smaller total surface area
than that of a fully perturbed surface that is deformed by large
wavelength instabilities.  However, in the MHD case 
the presence of external magnetic
fields may actually increase the drag on the bubble, and hence its
lifting effect, because of the long range nature of the fields.  A
self consistent MHD calculation is required to determine the magnitude
of this effect. 

Additional support for magnetic stabilization comes from the two-dimensional
MHD calculations of Br\"{u}ggen and Kaiser (2001). These calculations
consider a scale much larger than that appropriate for the relic
radio bubbles, with initial configurations extending from $200$ to
$400$ kpc and with source energies more appropriate for FR-II
radio sources than for the FR-I objects considered here.  In addition
these calculations do not consider the effects of the ambient magnetic
field in the ICM, but when a circumferential field is placed
around the radio bubble "by hand", the resulting geometry is similar
to that seen in the relic radio bubbles.  Moreover, this configuration 
shows no signs of disruption or mixing with the ICM.  It is important  
to note that this calculation of Br\"{u}ggen and Kaiser employs a 
magnetic field strength that has an energy density of 10 percent
that of the thermal energy in the ambient ICM (their "weak field" case).
While this field is dynamically unimportant, it is stronger than the
cluster magnetic fields used here.  However, the results are clearly
indicative of the effects of tangential magnetic fields around a
radio source bubble. 
The obvious
next step is a self-consistent three dimensional MHD calculation that
includes the ambient magnetic field in the intracluster medium, and
this is underway.  However, the above calculations for the onset of
disruptive instabilities in the presence of the magnetic field in
the ICM show that the transfer of energy from the radio bubbles to
the intracluster medium may not be prompt, and that the effectiveness
of such radio sources in reheating cooling flows may be less than
originally suggested.  

The final stage of transfer of energy from
a radio relic to the ICM takes place through the turbulent dissolution
of the radio source and its ultimate transfer into heat.  The time scale
for this very non-linear process is calculated next.

\subsection{Turbulent Mixing of Radio source Debris and ICM Heating}

Once a radio bubble has been disrupted by the Rayleigh-Taylor
instability, it develops large substructures which then break
down into ever smaller eddies and cells.  This process can be seen
in the high resolution 2-D hydrodynamic simulations of Br\"{u}ggen
\& Kaiser (2002), and it ultimately results in turbulent flow.  The
current numerical simulations do not follow the evolution of the
flow into this regime (Br\"{u}ggen et al. 2002).   
Though
such flows contain very fine structure, they do not actually heat
the ambient ICM until the turbulent cascade has proceeded down to
scales corresponding to the dissipation region.  A question of relevance
to the reheating of the ICM is the time scale for this process to
occur.  Using the wavelengths of the initial instability found above,
it is possible to calculate the development of fully non-linear MHD
turbulence in three dimensions (e.g., Orszag 1970, De Young 1992). 
The method employed solves the time dependent MHD equations in their
non-linear form, and the use of Fourier transformed second order
moments of velocity and magnetic field allows a calculation of 
the time evolution of both kinetic and magnetic energy over many orders
of magnitude in spatial resolution.  A wide variety of initial and
boundary conditions can be accommodated, and details of the method of
calculation, including the energy transfer among different wavenumbers
and the closure relations and dissipation scales, can be found in the
above references.  

The calculation here begins with continuous injection of energy at a 
single wavelength, which is the wavelength of the most unstable and
fastest growing mode calculated in Sect 2.1. Energy injected
at this scale quickly causes perturbations of the flow at smaller
scales, and this causes the development of a turbulent cascade of
energy to ever smaller scales or larger wavenumbers.  The multiple
interactions among turbulent eddies on different scales results in
the creation of a steady state turbulent flow with a loss free
propagation of energy through an inertial range into a dissipation
scale.  The calculation assumes that the magnetic field is 
homogeneous and weakly isotropic (allows helicity), since the 
non-linear stage of the R-T instability, once established, will
disrupt the original tangential field and advect it with the large
scale eddies as they rotate.  Equipartition of the turbulent magnetic
field with the kinetic energy will occur only on the smallest scales.
If for any reason the large scale tangential field persists, it will
inhibit the turbulent cascade and increase the time required for the
turbulent energy to be injected as heat into the intracluster medium.
Once an equilibrium state has been established, heat is being injected
into the ICM at the same rate as energy is being extracted from the
dissolving radio relic.  Figure 1 shows the results of such a
calculation in dimensionless form.  
The time development of the turbulence spectrum from
an initial delta function injection of energy at $k=1$
is clearly seen, and it is also clear that
an equilibrium spectrum is obtained after about $10$ large scale
eddy turnover times.  (Figure 1 shows the spectrum of turbulent
kinetic energy; the spectrum of the turbulent magnetic energy is
very similar for wavenumbers above $k=10$ where the magnetic
and kinetic energy quickly come into equipartition.) 

\begin{figure}
\includegraphics[width=85mm]{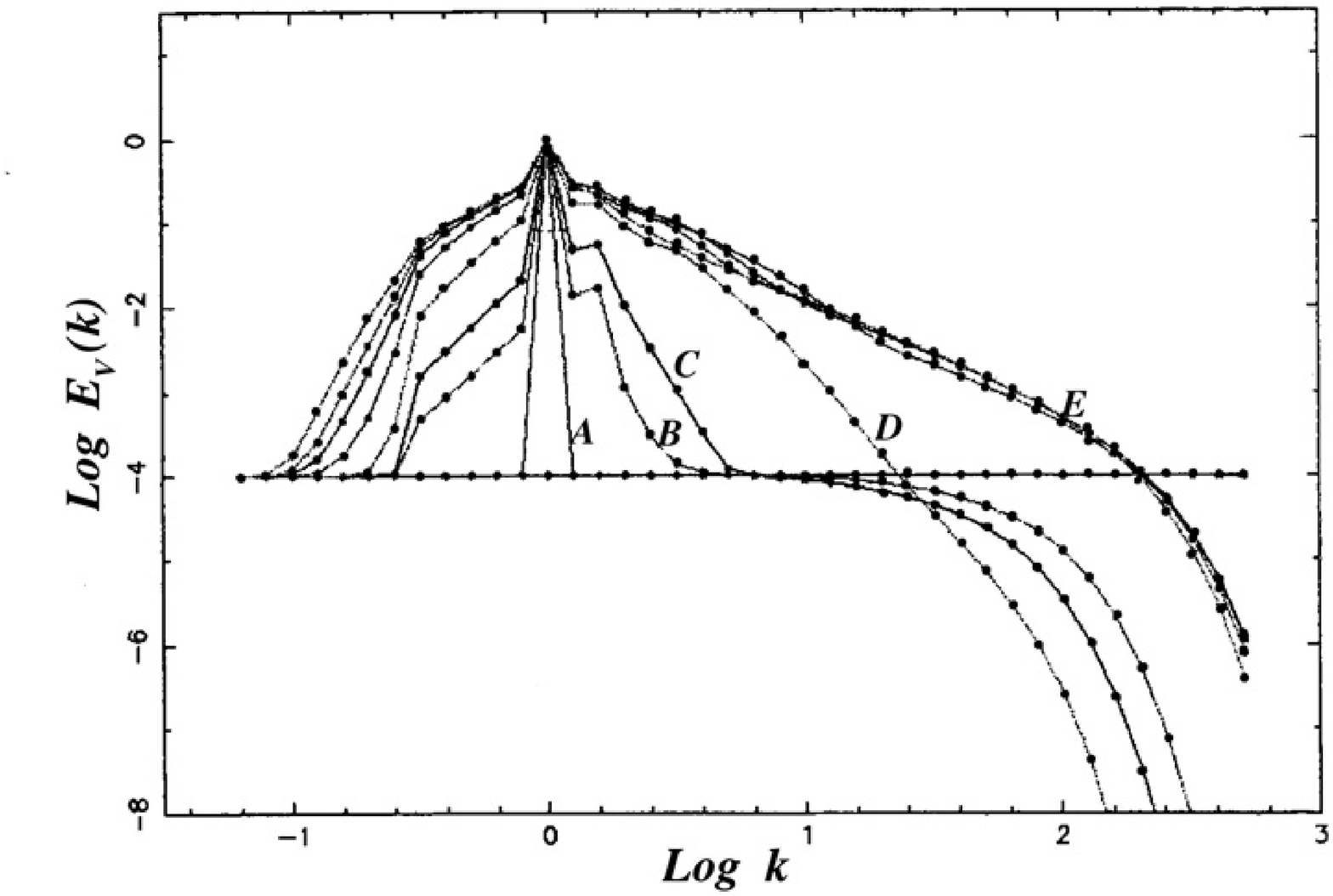}
 \caption{Turbulent kinetic energy per unit mass (in scaled units) as
  a function of wavenumber for various time in units of the eddy
  turnover time at $k=1$. A: initial injection spectrum; B: $t=1$;
  C: $t=2$; D: $t=5$; E: curves for $t=10, 15, 20 \& 30$.}
%  \label{sample-figure}
\end{figure}

The connection to the relic radio bubbles is
made by noting that the shortest wavelength for onset of the Rayleigh-Taylor
instability is $\simeq 15$ kpc, and that the wavelength for the 
fastest growing mode is $\sqrt{3}$ times this.  Using this as the
scale of the large eddy injection region, and using an eddy rotation
speed comparable to that of the speed of the rising radio bubbles
from the hydrodynamic simulations 
($300$ km s$^{-1}$)
gives 
$$
\tau_{edd} = \lambda_{edd}/v_{edd} = 4.4 \times 10^{7} yr.
$$
This provides the scaling factor for the scale free result shown
in Figure 1, and it 
then implies that once the instability begins, the actual heating
of the ICM commences about $4 \times 10^{8}$ years later.  
The calculation shown in Figure 1 does not extend to the actual
dissipation range, which lies at much smaller scales than considered in
the figure.  However,
extension of the turbulent spectrum to 
the dissipation range will occur on times much smaller than the
time required to establish equilibrium over the range shown due
to the much more rapid turnover time of the small scale turbulent
cells.
Hence this calculation, when coupled with the calculation for the
onset of the Rayleigh-Taylor instability in the presence of the
displaced ICM magnetic field, shows that the transformation of radio source
energy into heat input to the ICM may not take place until about
$5 \times 10^{8}$ years after the initial formation of the buoyant
radio source bubble. This is a time comparable to the cooling time
in the central regions of rich clusters (e.g., Fabian 1994), and
hence the overall energy balance between cooling of the ICM and
possible reheating due to the energy injected from radio sources
may need to be reconsidered.  

As mentioned at the end of Section 1, once
the energy is placed into the ICM as heat, it must be distributed
throughout the ICM if it is to have a global effect on the cooling
flows in clusters, and 
Br\"{u}ggen \& Kaiser (2002) estimated this effect via an averaging
calculation.  This global dissipation and mixing will take additional
time, presumably comparable to the orbital periods of the galaxies
in the cluster in the absence of large scale ICM motions due to 
subcluster mergers.  However, it is possible that local injection of
heat can have a significant effect on the cooling flow phenomenon
because local injection of heat can result in convective turnover being
induced in the ICM on scales comparable to that of the relic and now
virtually destroyed radio bubble.  This convective activity can serve
to mix hotter and cooler regions of the ICM, and if multiple events
such as this were to occur, an overall effect on the cooling flow could
result.  The timescale for this to take place, together with its spatial
extent, must be determined by a numerical calculation,
but $5 \times 10^{8}$ years seems an appropriate lower limit, since
the convection begins as soon as heating commences.

\subsection{Relation to Other Radio "Relics" in Clusters}

Diffuse radio emission from clusters of galaxies has been observed
in many clusters (e.g., Carilli \& Taylor 2002), and in some cases
the radio morphology of this emission is suggestive of that seen
in FR-I or FR-II radio sources, even though a parent galaxy cannot
be clearly identified.  Recent high resolution observations of 
four of these possible "relics" (Slee et al. 2001) show very
extended, diffuse and filamentary morphologies.  This appearance
is very different from the relic radio bubbles considered here,
and the sources observed by Slee et al. do not have clear and
unambiguous identifications of the parent galaxies.  The radio
morphologies are suggestive of much later stages of evolution of
radio relics, possibly under the influence of large-scale shearing
motions in the intracluster medium.  Kaiser \& Cotter (2002) have
modeled these objects as relics of FR-II radio sources.  Though
FR-II objects are not commonly seen in rich clusters, if these are
in fact FR-II remnants then their evolutionary tracks may be different
than those of the lower powered relic radio bubbles considered here.
In any case, it is interesting that even these sources do not appear
to have been completely assimilated into the ambient ICM, even at
the ages of greater than $10^{8}$ years considered by Kaiser \& Cotter.
This suggests again that the magnetic fields in both the ICM and
in the radio relics naturally inhibit the complete mixing of the
radio source with its environment, at least at some level. 
Calculation of the transfer of energy from the radio source to the
ambient ICM for these large sources and at late times would be of
interest in relation to cluster cooling flows.

\section{Implications for the Particle Content of Radio Sources}

A long standing issue in understanding extragalactic radio sources
is that of the nature of the energetic particle populations contained
within these objects (e.g., De Young 2002).  While relativistic
electrons are clearly present, the nature of the neutralizing particles,
whether protons or positrons, is still under debate.  The relic
radio bubbles appear to be in pressure equilibrium with the hot ICM,
yet they do not emit x-rays at the same intensity or energy as does
the ambient medium.  The question then arises as to the nature of the
material in the bubble that contributes the required energy density;
it could be magnetic field, hot protons, or very hot positrons. 
Hot positrons could arise from the very low
energy tail of a power-law distribution of electrons and positrons. 
If so, care has to be taken in choosing the low energy cutoff of such
a distribution; it must 
supply the required pressure, but it must not be such that the
electrons and positrons, themselves subject to Bremsstrahlung, will
emit x-rays above the observed limits.  For example, if a mildly
relativistic electron-positron population is used with 
$\gamma \simeq 1$, as has been suggested for some compact 
electron-positron jets (e.g., Celotti \& Fabian 1993),
then pressure equilibrium with the 
surrounding ICM at $p \simeq 10^{-10}$ dyne cm$^{-2}$ will give
rise to a Bremsstrahlung emissivity per unit volume from both
electrons and positrons that is roughly 10 percent of the 
radiation from the ambient ICM.  This emission should lie within
the sensitivity limits of the Chandra Observatory HRC but at the
upper end of its energy bandpass. 

An insight into this issue may arise from the possible need
to reaccelerate the synchrotron emitting
electron populations in the relic radio bubbles.  The dynamic lifetimes
are in excess of the radiative lifetimes (e.g., McNamara et al. 2001), and
the requirement for reacceleration also emerges from the calculation
of synchrotron radiation expected from the hydrodynamical simulations
(Br\"{u}ggen et al. 2002).  
Under the assumption that
reacceleration is the relevant process, and if 
such reacceleration is to take place
through internal shocks, Fermi acceleration, or turbulent acceleration
via stochastic processes, then the fluid within the bubble must be
able to sustain such motions and density inhomogeneities for
a time long enough to provide significant reacceleration.  If this fluid
is an electron-positron gas, and if it provides enough internal energy
to be in pressure balance with the hot external ICM, then this fluid
is mildly relativistic.  Such a hot gas has very
high sound speeds, and as a result it will rapidly damp any
discontinuities that produce shocks and will also not sustain large
density or pressure inhomogeneities for significant times; i.e., for
times in excess of many sound crossing times across the inhomogeneities.
The exception to this is if significant cooling can be made to take
place in density inhomogeneities, but in that case such areas would
be extremely bright in x-rays.  Hence it seems less likely that an
electron-positron gas can both provide pressure equilibrium in the
relic radio bubbles and also sustain the needed reacceleration of the
radio emitting electron population.  This is not the case for a
electron-proton gas, which can be hot but not relativistic ($T \sim
10^{10}$ K) and rarefied ($n \sim 10^{-4}$).  This fluid will have low
x-ray emissivity in the observed pass bands and will be in pressure
equilibrium with the surrounding ICM.  In this context it is interesting
to note that thermalization of a mildly relativistic jet ($ v_{j} \approx 
0.03c$) through a termination shock at the "working surface" will
result in a temperature of $10^{12}$ K.  Would such reacceleration
processes produce observable inhomogeneities in the radio emission from
the relic bubbles?  As mentioned above, the emission seems homogeneous
within the constraints of the low frequency observations.
Inhomogeneities in emission would be expected if very large scale
shock structures, comparable in size to the radio bubble, were present,
but such structures would not be expected in an aged radio relic.  A
much more likely process is turbulent reacceleration, and this would
occur more uniformly throughout the radio source volume on a "fine
grained" scale.  Thus it would be unlikely to produce large scale
intensity variations in the radio emission.
(An alternative to reacceleration is that of
"storage" of energetic particles in regions of low magnetic field, with
radiative losses taking place only in high field regions in an
inhomogeneous magnetic field distribution.  This process has been
suggested as a mechanism for larger FR-II and FR-I radio sources
(e.g., Eilek, Melrose \& Walker 1997, Tribble 1994), and though it 
cannot be ruled out here, the radio emission from the relic radio bubbles does
not display inhomogeneities or filamentary structures on the scale
resolved by current radio data.)   

If this reacceleration process is taking place as described, 
it implies that the electron-proton gas was present
in the radio source at early times, since the very existence of the
well defined and undisrupted relic radio bubbles implies that there
has been no significant mixing with the ICM, 
and that magnetic fields have kept the
internal and external fluids separated from the time of creation of
the radio source cavity.  This would imply that the electron-proton
gas was either ejected from the nucleus of the parent galaxy or that
the protons were entrained into the outflowing jet as it passed through
the parent galaxy's interstellar medium.  If magnetic isolation also
exists in the outwardly moving jet (which is more difficult to 
accomplish), then this would be evidence for an electron-proton gas
being present in the jet very close to its point of origin.

\section{Conclusions}

Recent observations of relic radio "bubbles" in the central regions
of rich clusters imply that the lifetime of these objects is longer
than that obtained from calculations of bubble disruption via the
onset and development of hydrodynamic Rayleigh-Taylor and Kelvin-Helmholtz
instabilities.  Inclusion of the effects of the magnetic field in
the ambient intracluster medium indicates that the creation of a
tangential field on the surface of these bubbles as they expand can 
stabilize them against such instabilities for times comparable to
their current lifetime.  Calculation of the time of onset of instability
under these conditions, together with calculation of the time required
for the energy in the radio source to be dissipated as heat in the
ICM, suggests that significant reheating of the ICM by radio sources
may not commence until about $5 \times 10^{8}$ years after the creation
of the radio source.  As this time is comparable to the cooling times
in the centers of rich clusters, the role of radio sources as a means
of reheating cooling flows may be more complex than earlier thought.
Finally, the existence of relic radio bubbles, together with their
likely pressure equilibrium with the surrounding intracluster medium,
can be used to argue for the presence of a hot gas interior to the
bubbles that is composed primarily of electrons and protons rather
than electrons and positrons.  The isolating effects of magnetic
fields interior and exterior to the radio bubbles then suggests that
the particle content of the original radio jet as it emerged from the parent
galaxy was primarily electrons and protons.

\vspace{.25in}

I thank Brian McNamara for many useful discussions and a
referee for helpful comments.

% \bsp % ``This paper has been produced using the ...''

\label{lastpage}

\end{document}